\documentclass[prd,aps,preprintnumbers,showpacs,10pt]{revtex4}
\usepackage{epsfig}
\newcommand{\be}{\begin{equation}}
\newcommand{\ee}{\end{equation}}
\newcommand{\bea}{\begin{eqnarray}}
\newcommand{\eea}{\end{eqnarray}}
\newcommand{\nn}{\nonumber}
\newcommand{\dd}{\displaystyle}

\def\slash#1{\setbox0=\hbox{$#1$}#1\hskip-\wd0\dimen0=5pt\advance
\dimen0 by-\ht0\advance\dimen0 by\dp0\lower0.5\dimen0\hbox
to\wd0{\hss\sl/\/\hss}} 
\begin{document}
\begin{titlepage}

\preprint{BARI-TH 513/05}

\title{$B\to K_1\,\gamma $ and tests of factorization for two-body  non
  leptonic $B$ decays with axial-vector mesons}

\author{\textbf{G. Nardulli}}
\affiliation{Dipartimento di Fisica dell'Universit{\`a} di Bari, Italy\\
Istituto Nazionale di Fisica Nucleare, Sezione di Bari, Italy}
\author{\textbf{T. N. Pham}}
\affiliation{Centre de Physique Th{\'e}orique,\\
 Centre National de la Recherche Scientifique, UMR 7644, \\
{\'E}cole Polytechnique, 91128 Palaiseau Cedex, France}

\begin{abstract}
The large branching ratio for $B \to K_{1}\,\gamma$ recently
measured at Belle implies a large  $B \to K_{1}$ transition form
factor and large branching ratios for  non leptonic $B$ decays
involving an axial-vector meson. In this paper we present an
analysis of two-body $B$ decays with an axial-vector meson in the
final state using naive factorization and the $B \to K_{1}$ form
factors obtained from the measured radiative decays. We find that
the predicted $B \to J/\psi\,K_{1}$ branching ratio is  in
agreement with experiment. We also suggest that the decay rates of
$B \to K_{1}\,\pi$, $B \to a_{1}\, K$ and $B \to b_{1}\, K$ could
be used to test the factorization ansatz.

\end{abstract}


\maketitle
\end{titlepage}
\setcounter{page}{1}
\newpage
\section{Introduction}
Our analysis is based on the recent announcement from the Belle
collaboration concerning  the first measurement of the branching
ratio ${\cal B}$ for $B$ decay into $K_1(1270)\,\gamma$
\cite{Yang:2004as}:
\begin{equation}
{\cal B}(B^+\to K_1^+(1270)\gamma)= (4.28\pm0.94\pm0.43)\times
10^{-5}~, \label{1270}
\end{equation}together with an upper bound on $K_1(1400)$:
\begin{equation}
{\cal B}(B^+\to K_1^+(1400)\gamma)< 1.44\times 10^{-5}~({\rm at
~90\% ~C.L.}) . \label{1400}
\end{equation}
These results should be compared to $B\to K^*\gamma$. The decay
fractions measured by CLEO \cite{Coan:1999kh}, BaBar
\cite{Aubert:2001me} and  Belle \cite{Nakao:2004th} Collaborations
result in average branching ratios
 $
{\cal B}(B^0\to K^{*0}\gamma) =(4.17\pm0.23)\times 10^{-5}$ and
$ {\cal B}(B^+\to K^{*+}\gamma) =(4.18\pm0.32)\times 10^{-5}$ .

The large measured ${\cal B}(B^+\to K_1^+(1270)\gamma )$ is a
surprise since recent calculations
\cite{Safir:2001cd,Lee:2004ju,Kwon:2004ri,Cheng:2004yj} predict a
branching ratio smaller than the measured value by a factor
$\approx 4$, though a previous calculation \cite{Ali:1992zd} gives
a larger branching ratio, in the range $(1 - 4)\times 10^{-5}$ ,
not too far from the measured value. However the small tensor $B
\to K_{1}$ form factor for the radiative decays $B \to
K_{1}\,\gamma$ obtained by these recent calculations implies also
a tiny  branching ratio for non leptonic two-body $B$ decays with
axial-vector meson in the final state. Therefore one would expect
a small branching ratio for $B \to J/\psi\,K_{1}$. This is in
contrast with the large measured value \cite{Abe:2001wa} for this
decay. This value is comparable with the $B \to J/\psi\,K^{*}$
branching ratio, which implies that the  form factors for the
transitions $B \to K_{1}(1270)$ and  $B \to K^{*}$ should be
similar in size in order to explain the large branching ratios for
both the radiative and the non leptonic $B\to K_{1}(1270)$ decays.
 The aim of the present letter is to present arguments to
 show that this is  indeed the case.
  We employ naive factorization and the heavy quark
symmetry to relate the tensor form factor of the radiative
transition to the  form factors that describe non-leptonic decays.
From the measured radiative decay rates as well as recent data on
branching ratios and polarizations for $B \to J/\psi K^{*}$
decays, we find that the predicted ${\cal B}(B \to J/\psi
K_{1}(1270))$ agrees with the experimental results.

From the data (\ref{1270}), (\ref{1400}) we also derive some
straightforward  predictions for a few  non leptonic decay
channels involving light strange or non-strange axial-vectors in
the final state. This can be achieved by making use
 only of naive factorization and relations obtained from the
Heavy Quark Effective Theory (HQET)
\cite{Isgur:1989ed,Isgur:1989vq,Neubert:1993mb}. Our approach is
therefore in the spirit of the chiral effective theory for heavy
mesons, see e.g.
\cite{Wise:1992hn,Casalbuoni:1992gi,Casalbuoni:1993dx,Cheng:1992xi,Cheng:1994kp},
and for a review \cite{Casalbuoni:1997pg}, where a similar
approach was used to relate a number of decay channels of heavy
mesons using the approximate symmetries of HQET.

\section{$B\to K_1$
radiative decays and the mixing angle}
 The $K_1(1270)$ and $K_1(1400)$ are strange
axial-vector resulting from a mixing of $^3P_1$ and $^1P_1$
states. Following PDG \cite{PDBook}, we denote by $K_{1A}$ and
$K_{1B}$ the $^3P_1$ and $^1P_1$ states of $K_1$. Thus we have
 \bea \label{mix}
 K_1(1270)&=&K_{1A} \sin\theta+K_{1B}\cos\theta,
 \cr
 K_1(1400)&=&K_{1A} \cos\theta-K_{1B}\sin\theta\ .\label{eqmix}
 \eea
The mixing angle $\theta$ has been determined up to a fourfold
ambiguity, see  \cite{Cheng:2004yj} and, previously,
\cite{Suzuki:1993yc}. The masses of $K_{1A}$ and $K_{1B}$, can be
determined by the relations \cite{Suzuki:1993yc}\bea
m_{K_{1A}}^2&=&m_{K_1(1270)}^2+m_{K_1(1400)}^2-m_{K_{1B}}^2\ ,\cr
2m_{K_{1B}}^2&=&m_{b_1(1235)}^2+m_{h_1(1380)}^2\ .\eea  $K_{1B}$
belongs to the same  nonet as the states  $b_1(1235)$, $h_1(1170)$
and $h_1(1380)$; $K_{1A}$, $a_1(1260)$, $f_1(1285)$ and
$f_1(1400)$ are also in the same nonet, different from the
previous one. Besides (\ref{eqmix}) we also have \be\cos
2\theta\,=\, \frac{m^2_{K_{1B}}-m^2_{K_{1A}}}{m_{K_1(1270)}^2
-m^2_{K_1(1400)}}\ .\label{mix2}\ee Using (\ref{eqmix}) and
(\ref{mix2}) and restricting to $0<\theta<90^0$ we get only two
solutions \cite{Suzuki:1993yc} :\bea {\rm Sol.
[a]}:\hskip1cm\theta&=&32^o\ ,\hskip.5cm
(m_{K_{1B}},\,m_{K_{1A}})=(1310,\ 1367)\ {\rm MeV}\ ,\cr {\rm Sol.
[b]}:\hskip1cm\theta&=&58^o\ ,\hskip.5cm
(m_{K_{1B}},\,m_{K_{1A}})=(1367,\ 1310)\ {\rm MeV}\ .\eea

These results give a clue for understanding the Belle results. In
fact, for any reasonable computational scheme the form factors
$T_1(0)$  that determine the radiative decays $B\to K_{1A}\gamma$
and $B\to K_{1B}\gamma$ should be almost identical. This is
confirmed by the dynamical calculation of Ref. \cite{Cheng:2004yj}
that gives for this ratio \be \frac{T_1^{B\to K_{1B}
}(0)}{T_1^{B\to K_{1A}}(0)}=1.2 \label{cheng}\ ,\ee where the form
factor is defined by \bea\langle K_1(p^\prime,\,\epsilon)|{\bar
s}\sigma_{\mu\nu}(1+\gamma_5) q^\nu b|
B(p)\rangle\,&=&\,i\epsilon_{\mu\nu\rho\sigma}\epsilon^{*\,\nu}p^\rho
p^{\prime\,\sigma}2\,T_1(q^2)\,+\,[\epsilon_\mu^{*}(m^2_B-m^2_{K_1})\,-\,(\epsilon^*\cdot
q)(p+p^\prime)_\mu]T_2(q^2)\cr&+&[(\epsilon^{*}\cdot
q)q_\mu\,-\,\frac{q^2}{m^2_B-m^2_{K_1}}(p+p^\prime)_\mu]T_3(q^2) \
. \label{t1} \eea Here $T_1(0)=T_2(0)$, while $T_3$ does not
contribute to the radiative decay. A similar definition holds also
for $B\to K^*$. We note that, from experiment, \be y\equiv
\frac{T_1^{B\to K_{1}(1270)}(0)}{T_1^{B\to
K^{*}}(0)}=\sqrt{\left(\frac{m_B^{2}-m^2_{K^*}}{m_B^{2}-m^2_{K_1}}\right)^3\frac{{\cal
B}(B\to K_1(1270) \gamma)}{{\cal B}(B\to K^*\gamma)}}\approx 1.06\
.\label{y}\ee

As to $K_1(1400)$ we get \be \frac{{\cal B}(B\to K_1(1400)
\gamma)}{{\cal B}(B\to K_1(1270) \gamma)}=
\left(\frac{m^2_B-m^2_{K_{1}}(1400)}{m^2_B-m^2_{K_1}(1270)}\right)^3
\Big| \frac{T_1^{B\to K_{1A} }(0)-\,\tan\theta\,\, T_1^{B\to
K_{1B} }(0)}{T_1^{B\to K_{1B} }(0)+\tan\theta\,\, T_1^{B\to K_{1A}
}(0)}\Big|^2\ .\ee Assuming the value (\ref{cheng}) we can
predict, from the Belle result (\ref{1270}) the value for ${\cal
B}(B\to K_1(1400) \gamma)$. The result is in Table \ref{tab:1}.
Both the solutions obtained are in agreement with the upper limit
(\ref{1400}).

\section{$K_1$ leptonic decay constant} $K_1$
leptonic decay constant can be derived from $\tau$ decays. Let us
denote by ${\cal A}$ a generic axial-vector meson, i.e one of the
following states: $K_{1A}\,,K_{1B}\,,a_1\,,b_1$. We also denote by
$P\,,P^{(\prime)}$ the pseudoscalar mesons, and we use the
following definition for the matrix elements of weak currents:
\be \label{leptonic} \langle 0\, |A_\mu|\, P(p)\rangle = i\, f_P\,
p_\mu ~,~~~~~~~~~~~~\langle {\cal
A}(\varepsilon,p)|A_\mu|\,0\,\rangle =
 f_{\cal A}\, m_{\cal A}\, \varepsilon^\ast_\mu\ .
\ee From the $\tau\to K_1$ data \cite{PDBook} we get\be
f_{K_1(1270)}=171 \ {\rm MeV}\,,\hskip1cm f_{K_1(1400)}=126 \ {\rm
MeV}\label{f}\ee Using the mixing angle we derived and $SU(3)$
symmetry we get \bea {\rm Sol.
[a]}\hskip.3cm(\theta&=&32^0):\hskip.5cm
(f_{b_{1}},\,f_{a_{1}})=(74,\ 215)\ {\rm MeV}\ ,\cr {\rm Sol.
[b]}\hskip.3cm(\theta&=&58^0):\hskip.5cm
(f_{b_{1}},\,f_{a_{1}})=(-28,\ 223)\ {\rm MeV}\ .\eea We note that
these values might be useful to compute weak decays with non
strange axial vector in the final state.
\section{$B\to K_1 J/\psi$\label{SECpsi}}

 For the decay $B\to K^* J/\psi$ and $B\to K_1 J/\psi$ we have
 the experimental result reported in Table \ref{tab:1} and we may ask  if they
 are compatible with the Belle
 result (\ref{1270}).

We use a simple scaling relations, based on HQET, which allows to
relate the form factors for the transition $B\to K^*$ {\it via}
V-A current to those describing transitions by a tensor current.
At large $q^2$ it relates the $A(q^2)$ and $V_1(q^2)$ form factors
defined by \bea <{\cal A}(\epsilon,p')|V^\mu -A^\mu|P(p)>& =&\,
+\, i (m_P+m_{\cal A}) \epsilon^{*\mu}
 V_1 (q^2) - i \frac{(\epsilon^* \cdot q)}{m_P+m_{\cal A}} (p+p')^\mu V_2
(q^2)\nn\\
& -& i (\epsilon^* \cdot q)\, \frac{2 m_{\cal A}}{q^2} q^\mu\left[
V_3 (q^2)- V_0 (q^2)\right]\,-\,
 \frac {2 A(q^2)}
{m_P+m_{\cal A}} \epsilon^{\mu \nu \alpha \beta}\epsilon^*_{\nu}
p_{\alpha} p'_{\beta} \label{v1} \eea where \be V_3 (q^2)=\frac
{m_{\cal A}-m_P}{2 m_{\cal A}} \, V_2(q^2)\, + \,\frac {m_{\cal
A}+m_P} {2m_{\cal A}} \, V_1(q^2) \label{v2}\ee and
$V_3(0)=V_0(0)$, with $T_1(q^2)$ in (\ref{t1}) as follows \be
T_1(q^2)=\frac{q^2+m^2_B-m^2_{K_1}}{2m_B}\cdot\frac{A(q^2)}{m_B+m_{K_1}}
\,-\,\frac{m_B+m_{K_1}}{2m_B}\,V_1(q^2)\ . \ee Moreover we assume
that the effect of substituting $K^*$ with $K_1$ is identical in
the radiative and in the non leptonic decay, in other words that
each form factor  for the $B\to K_1$ transition is given by the
corresponding form factor for $B\to K^*$ multiplied by the same
factor $y$, once the change of parity between the two strange
mesons is taken into account. On this basis we predict
\be\frac{{\cal B}(B\to K_1(1270) J/\psi)}{{\cal B}(B\to K^*
J/\psi)}\frac{{\cal B}(B\to K^*\gamma)}{{\cal B}(B\to K_1(1270)
\gamma)}=\frac{p_{K_1}}{p_{K^*}}\left(\frac{m_B^{2}-m^2_{K^*}}{m_B^{2}-m^2_{K_1}}\right)^3
\left(x_\parallel+x_\bot\,\frac{p^2_{K_1}}{p^2_{K^*}}+x_L\cdot\frac{m^2_{K^*}}{m^2_{K_1}}\right)
\ee Here (we use the BaBar data \cite{Aubert:2001pe})
 \bea
 x_\parallel &=&
 \frac{\Gamma_\parallel(B\to K^*J/\psi)}{\Gamma(B\to K^* J/\psi)}=
0.24\pm 0.04
  \cr
  x_\bot &=&
\frac{\Gamma_\bot(B\to K^* J/\psi)}{\Gamma(B\to K^*
J/\psi)}=0.16\pm 0.03\cr
 x_L &=& \frac{\Gamma_L(B\to K^*
J/\psi)}{\Gamma(B\to K^* J/\psi)}= 0.60\pm 0.04\label{psi}\eea
while $p_{K^*}$ (resp. $p_{K_1}$) is the c.m momentum of $K^*$
(resp. $K_1$) for the nonleptonic decay $B\to K^*  J/\psi$ (resp.
$B\to K_1 J/\psi$).

The r.h.s of eq. (\ref{psi}) has the numerical value ${\rm
r.h.s.}=0.64$, while \be{\rm l.h.s.}= \ \  \left\{ ^{\dd 0.94
\hskip.4cm {\rm (neutral~~mode)}}_{\dd  1.30\hskip.4cm {\rm
(charged~~ mode)}}\right. \ee with experimental uncertainties of
around 50\%. Thus we see that the experimental results for $B\to
K_1(1270)\gamma$ and $B\to K_1(1270)\,J/\psi$ are compatible
within the errors. We report in Table \ref{tab:1} our prediction.
Similar arguments apply to the decay $B\to K_1(1400)\,J/\psi$.
Also these results can be found in Table \ref{tab:1}.
\section{$B\to K_1\pi$}
For $B\to K_1\pi$ decays, if $q_{K_1}$ and $q_{K^*}$ are
respectively the c.m. momenta of ${K_1}$ and ${K^*}$ in the
reactions $B\to K_1 \pi$ and $B\to K^{*} \pi$, one gets, using
factorization: \be\frac{{\cal B}(B^+\to K^0_1 \pi^+)}{{\cal
B}(B^+\to K^{*\,0} \pi^+)}\, =\, \frac{{\cal B}(B^0\to K^+_1
\pi^-)}{{\cal B}(B^0\to K^{*\,+} \pi^-)}\,= \,
\left(\frac{q_{K_1}}{q_{K^*}}\right)^3\,\frac{m^2_{K^*}}{m^2_{K_1}
}\,\left(\frac { F_1^ {B\to\pi} (m^2_{K_1}) \,f_{K_1} \,m_{K_1} }
{ F_1^ {B\to\pi} (m^2_{K^*}) f_{K^*} m_{K^*}
}\right)^2\label{pred0}\ .\ee Here we use the form factor $F_1$
defined by \be \langle P^\prime(p^\prime)|V_\mu|P(p)\rangle =
F_1(q^2)
\left[(p_\mu+p^\prime_\mu)-\frac{m_P^2-m_{P^\prime}^2}{q^2}\,
q_\mu\right]+ F_0(q^2)\frac{m_P^2-m_{P^\prime}^2}{q^2}\, q_\mu\ee
and a simple pole formula, with a pole mass equal to $m_{B^*}$,
for the $q^2$ behavior of the $F_1^{B\to\pi}$ form factor. The
results obtained by (\ref{pred0}) are reported in Table
\ref{tab:1} and represent an interesting  test of factorization.
It is indeed quite possible that both $B\to K^*\pi$ and $B\to
K_1\pi$ decays take non-factorizable contributions from long
distance  operators formally suppressed in the $m_b$ limit, see
e.g. \cite{Isola:2003fh}, or power corrections in QCD
Factorization \cite{Pham:2004xw}. In this case the predictions of
the last four rows in Table \ref{tab:1} would get significant
violations, pointing to non-factorizable contributions to the
decay amplitude.

The  reactions with a $\pi^0$ in the final
state: $B^+\to K^+_1\pi^0$ and $B^+\to K^+_1\pi^0$ involve two
form factors $F_1$ and $V_0$ and different combinations of Wilson
coefficients and CKM matrix elements. As explained in the
introduction the main purpose of this letter is to pick up a few
decay channels involving light and strange axial-vector mesons in the
final state whose rates can be predicted using only the Belle
results \cite{Abe:2004kr}, $\tau$ decay rates and the
factorization hypothesis. On this basis we skip these channels
leaving a complete analysis to a future paper.

\section{$B\to {\cal A}_1 K$}

Also in this case we have some clear predictions based on
factorization:
 \bea
 %%%%%%%%%%%%%%1%%%%%%%%%%
 \frac{{\cal B}(B^+\to a_1^+ K^0)_{\rm fact.}}{{\cal
B}(B^+\to \rho^+ K^{0})_{\rm fact.}}
%&=&
%\frac{{\cal B}(B^0\to a_1^-
%K^+)}{{\cal B}(B^0\to \rho^-
%K^{+})}=
%\left(\frac{q_{a_1}}{q_{\rho}}\right)^3\,\left(\frac { V_0^
%{B\to\,a_1} (m^2_{K}) } { A_0^{B\to\rho} (m^2_{K}) }\right)^2\,R_+
%\cr&& \cr
&\approx
&\left(\frac{q_{a_1}}{q_{\rho}}\right)^3\,\left(\sin\theta\,\frac
{ V_0^ {B\to\,K_1(1270)} (m^2_{K}) }{ A_0^{B\to\rho}
(m^2_{K})}+\cos\theta\,\frac { V_0^ {B\to\,K_1(1400)} (m^2_{K}) }{
A_0^{B\to\rho} (m^2_{K})}\right)^2\,R_+\label{a1}
\\ &&\cr
%%%%%%%%%%%%%%%%2%%%%%%%%%%%%%%
\frac{{\cal B}(B^+\to b_1^+ K^0)_{\rm fact.}}{{\cal B}(B^+\to
\rho^+ K^{0}_{\rm fact.})}
%&=&
%\frac{{\cal B}(B^0\to b_1^- K^+)}{{\cal B}(B^0\to
%\rho^-
%K^{+})}&=&
%\left(\frac{q_{b_1}}{q_{\rho}}\right)^3\,\left(\frac { V_0^
%{B\to\,b_1} (m^2_{K}) } { A_0^{B\to\rho}
%(m^2_{K})}\right)^2\,R_-\cr&& \cr
&\approx&\left(\frac{q_{b_1}}{q_{\rho}}\right)^3\,\left(\cos\theta\,\frac
{ V_0^ {B\to\,K_1(1270)} (m^2_{K}) }{ A_0^{B\to\rho}
(m^2_{K})}-\sin\theta\,\frac { V_0^ {B\to\,K_1(1400)} (m^2_{K}) }{
A_0^{B\to\rho} (m^2_{K})}\right)^2\,R_+ \label{a2}\\ && \cr
%\eea
% \bea%\frac{{\cal B}(B^+\to a_1^+ K^0)}{{\cal
%B}(B^+\to \rho^+ K^{0})}&=&
%%%%%%%%%%%%%%%%%%%%%3%%%%%%%%%%%%%%%%%
\frac{{\cal B}(B^0\to a_1^- K^+)_{\rm fact.}}{{\cal B}(B^0\to
\rho^- K^{+})_{\rm fact.}}
%&=&\left(\frac{q_{a_1}}{q_{\rho}}\right)^3\,\left(\frac {
%V_0^ {B\to\,a_1} (m^2_{K}) } { A_0^{B\to\rho} (m^2_{K})
%}\right)^2\,R_- \cr&& \cr
&\approx&\left(\frac{q_{a_1}}{q_{\rho}}\right)^3\,\left(\sin\theta\,\frac
{ V_0^ {B\to\,K_1(1270)} (m^2_{K}) }{ A_0^{B\to\rho}
(m^2_{K})}+\cos\theta\,\frac { V_0^ {B\to\,K_1(1400)} (m^2_{K}) }{
A_0^{B\to\rho} (m^2_{K})}\right)^2\,R_-\label{a3}
\\ &&\cr
%\frac{{\cal B}(B^+\to b_1^+ K^0)}{{\cal B}(B^+\to \rho^+
%K^{0})}&=&
%%%%%%%%%%%%%%%%%%%%%4%%%%%%%%%%%%%%%
\frac{{\cal B}(B^0\to b_1^- K^+)_{\rm fact.}}{{\cal B}(B^0\to
\rho^- K^{+})_{\rm fact.}}
%&=&\left(\frac{q_{b_1}}{q_{\rho}}\right)^3\,\left(\frac {
%V_0^ {B\to\,b_1} (m^2_{K}) } { A_0^{B\to\rho}
%(m^2_{K})}\right)^2\,R_-\cr&& \cr
&\approx
&\left(\frac{q_{b_1}}{q_{\rho}}\right)^3\,\left(\cos\theta\,\frac
{ V_0^ {B\to\,K_1(1270)} (m^2_{K}) }{ A_0^{B\to\rho}
(m^2_{K})}-\sin\theta\,\frac { V_0^ {B\to\,K_1(1400)} (m^2_{K}) }{
A_0^{B\to\rho} (m^2_{K})}\right)^2\,R_- \label{a4}\eea where the
subscript means that we consider only factorizable  contributions.
$V_0$ has been defined in (\ref{v1}) and, if $|V\rangle$ is a
vector meson state,\bea <{V}(\epsilon,p')|V^\mu -A^\mu|P(p)>& =&\,
-\, i (m_P+m_{V}) \epsilon^{*\mu}
 A_1 (q^2) + i \frac{(\epsilon^* \cdot q)}{m_P+m_{V}} (p+p')^\mu A_2
(q^2)\nn\\
& +& i (\epsilon^* \cdot q)\, \frac{2 m_{V}}{q^2} q^\mu\left[ A_3
(q^2)- A_0 (q^2)\right]\,+\,
 \frac {2 V(q^2)}
{m_P+m_{\cal A}} \epsilon^{\mu \nu \alpha \beta}\epsilon^*_{\nu}
p_{\alpha} p'_{\beta}\label{a44} \eea
and
\be
  A_3(q^2)
  =
  \frac{m_{V}-m_P}{2 m_{V}} A_2(q^2) + \frac {m_{V}+m_P} {2m_{V}} A_1(q^2)
  \label{v2bis}\ee
with  $A_3(0)=A_0(0)$; $q_{a_1}$ and $q_{b_1}$ are the c.m.
momenta of $a_1$ and $b_1$ respectively; the factors $R_\pm$ are
defined below.

It is a well known fact that factorization terms give small
contribution to the decay rates $B^+\to \rho^+ K^0$, $B^0\to\rho^-
K^+$; for example, for the $B^{0}\, \to \rho^{-} K^+$ channel, the
experimental result ${\cal B}(B^{0}\, \to \rho^{-} K^+)=(7.3\pm
1.8)\times 10^{-6}$ \cite{PDBook} is larger by one order of
magnitude than theoretical predictions based on factorization
\cite{Isola:2003fh}, \cite{Ali:1998eb}. This is mainly due to the
large cancellation between the penguin contributions appearing in
the denominator of the two factors $R_\pm$. These two factors
differ by 1 because of the different parity of the vector and
axial-vector mesons. The penguin operators $O_6$ and $O_8$
distinguish the two parities and therefore \bea
R_+&=&\left(\dd\frac{\dd
a_4-\frac{a_{10}}2+\frac{(2\,a_6-a_8)\,m^2_K}{(m_b-m_d)(m_s+m_d)}}{\dd
a_4-\frac{a_{10}}2-\frac{(2\,a_6-a_8)\,m^2_K}{(m_b+m_d)(m_s+m_d)}}\right)^2\
,
\\ &&\cr R_- &=&\left(\dd\frac{\dd
a_4+a_{10}+\frac{2(a_6+a_8)\,m^2_K}{(m_b-m_u)(m_s+m_u)}}{\dd
a_4+a_{10}-\frac{2(a_6+a_8)\,m^2_K}{(m_b+m_u)(m_s+m_u)}}\right)^2
\ .\eea For numerical evaluation of these coefficients we take
\cite{Buras:1998ra}: $ c_3=0.013,~c_4=-0.029,~c_5=0.009,
~c_6=-0.033,~c_7/\alpha=0.005,~c_8/\alpha=0.060,
~c_9/\alpha=-1.283,~c_{10}/\alpha=0.266$, with ${\dd a_i = c_i +
\frac{c_{i-1}}{3}}$ (i=even). The other two Wilson coefficients
$c_2=1.105,~c_1=-0.228$ are of no interest here. Moreover, for the
current quark masses we use the values $ m_b=4.6$\, GeV,
$m_u\,=\,4$\,MeV\, $m_d\,=\,8\,$MeV\,, $m_s\,=\,0.150$\,GeV.
 We get therefore
\be R_+\,\approx\,160\,,\hskip1cm R_-\,\approx \,80\ .\ee
 Following the same procedure of Section
\ref{SECpsi} we evaluate the ratio of form factors as follows. \be
\frac { V_0^{B\to\,K_1} (m^2_{K})
}{A_0^{B\to\rho}(m^2_{K})}\,\approx\,\frac { V_0^{B\to\,K_1} (0)
}{A_0^{B\to\rho}(0)}=\, y\, \frac{m_{K^*}}{m_{K_1}} \,
\frac{m_B+m_{K_1}-(m_B-m_{K_1}) \,z}{m_B+m_{K^*}-(m_B-m_{K^*})\,z}
\ee Here $y$ is defined, for $K_1(1270)$ by (\ref{y}); a similar
expression holds for $K_1(1400)$ and $y_{K_1(1400)}=0.14$
 for $\theta=32^o$ and $y_{K_1(1400)}=0.35$  for $\theta=58^o$ . The factor z is defined as \be
z=\frac{A_2^{B\to\rho}(0)}{A_1^{B\to\rho}(0)}\approx
\frac{A_2^{B\to K^*}(0)}{A_1^{B\to K^*}(0)}\ee We take the value
$z=0.93$ intermediate between the value $z=0.9$ predicted by light
cone sum rules \cite{Ball:2004rg} and $z=0.95$ given by the BWS
model \cite{Bauer:1986bm}. Although the phase space and the ratio
of form factors act as suppressing factors, the big enhancement
given by $R_\pm$ can produce very large predictions for the decays
$B\to a_1,\,b_1 K$. As a matter of fact we get for the four ratios
in eqns. (\ref{a1})-(\ref{a4}) results of the order $\approx
(59,\,76,\,29,\,38)$ for the solution $\theta=32^o$ and $\approx
(147,\,9,\,73,\,4)$ for the solution $\theta=58^o$. This means
that factorization terms give sizeable contributions to these
decays, and especially to $B\to a_1^+ K^0$ for both values of the
mixing angle. Our conclusion is that, in view of these results,
two-body nonleptonic $B$ decays with a kaon and a light
non-strange axial vector meson  in the final state represent
interesting decay channels with expected large branching ratios.
Significant experimental deviations from the the abovementioned
ratios would point to specific violations of the factorization
model.

It would be tempting to extend the present analysis to the case of
$B$ transitions to other orbitally excited $K$ mesons. For example
two decay modes with $K_2^{*}(1430)$ in the final state have been
measured: $B\to K_2^{*}(1430)\gamma$ and $B\to
K_2^{*}(1430)J/\psi$. The radiative transitions $B\to
K_2^{*}(1430)$ have been investigated by some authors. In Ref.
\cite{Veseli:1995bt} HQET is used and the strange quark is treated
as heavy, which is however a rather crude approximation. As a
result, these authors predict ${\cal B}(B\to
K_2^{*}(1430)\gamma)/{\cal B}(B\to K_1(1270)\gamma)=3$, which is
at odds with the data, though the predicted $B\to
K_{2}^{*}(1430)\gamma$ branching ratio is in agreement with
experiments. On the contrary in \cite{Ebert:2001en} the $s$ quark
is considered light. Also this relativistic quark model reproduces
correctly the $B\to K_2^{*}(1430)\gamma$ decay mode, but predicts
a too small branching ratio for $B\to K_1(1270)\gamma$. This again
brings up the problem with the
 small predicted radiative branching ratio involving $K_{1}(1270) $ state,
the motivation for the present work. There are also more recent
calculations \cite{Cheng:2004yj} with  results in agreement with
experiment for the $B\to K_{2}^{*}(1430)\gamma$ branching ratios,
obtained by various techniques such as light cone sum rules or
covariant relativistic quark models as given in the Table V of
\cite{Cheng:2004yj}. An analysis  of the $B\to
K_2^{*}(1430)J/\psi$ decay mode is performed in \cite{Kim:2003rz}.
These authors use the ISGW2 quark model \cite{Scora:1995ty},
\cite{Isgur:1988gb} and find results that are however sensitive to
the model-dependent form factors. Tests of factorization would be
therefore desirable also for these channels. However it must be
said  that the extension of the present study of $B\to K_1$
transitions to the $B\to K_2^{*}(1430)$ decay modes cannot be
immediate. To study $B\to K_1$ transitions we have used data from
$B\to K^*(892)$ transitions and made some further hypotheses,
based on the chiral similarity between $1^-$ and $1^+$ states (see
the discussion on the $B\to K_1 J/\psi$ channel presented above).
For $B\to K_2^{*}(1430)$ we are in a less favorable situation and
some further assumption has to be made. We plan to come back to
this issue in the future.

\vspace{0.8truecm} \textbf{Acknowledgements} One of us (G.N.)
thanks the Centre de Physique Th{\'e}orique,  CNRS, {\'E}cole
Polytechnique, Palaiseau, where this work was partly done,  for
kind hospitality.

\newpage
\begin{table}[ht!]
\caption{\label{tab:1} {\small  Theoretical branching ratios and
comparison with experiment; [a] and [b] refer to the two possible
values of the mixing angle between the $K_{1A}$ and $K_{1B}$
states, $\theta=32^o$ and $\theta=58^o$ respectively. The
experimental values for the processes in  lines 1, 2, 3, 5, 9 and
10 are used as inputs.}}
\begin{footnotesize}
\begin{center}
\begin{tabular}{|c|c|c|}
\hline
{\rm Process}  & ${\cal B}\,\, ({\rm theory})$ & {\rm exp.}\\
%%%%%%%%%%%
\hline

& \vspace{-0.18truecm} &\\
$  B^{+}\, \to K^{\ast +}\gamma$ &  {\rm input}&$(4.18\pm
0.31)\times 10^{-5}$\,\,
({\rm av. of \cite{Coan:1999kh,Aubert:2001me,Naka:2004}}) \\
& \vspace{-0.18truecm}& \\
\hline

$ B^0\, \to K^{\ast 0} \gamma$& {\rm input}&$(4.17\pm 0.23)\times
10^{-5}$\,\,
({\rm av. of \cite{Coan:1999kh,Aubert:2001me,Naka:2004}}) \\

& \vspace{-0.18truecm}& \\
\hline
& \vspace{-0.18truecm} &\\

$  B^+\, \to  K^{+}_1(1270) \gamma$ & {\rm input}&
$(4.28\pm0.94\pm 0.43)\times 10^{-5}$\,\, {\rm \cite{Abe:2004kr}}\\

& \vspace{-0.18truecm}& \\
\hline
& \vspace{-0.18truecm}& \\

& $7.7\times 10^{-7}$\hskip.5cm [a]&\\

$  B^{+}\, \to K^{+}_1(1400) \gamma$&
& $< 1.44\times 10^{-5}$\,\, \cite{Abe:2004kr}\\

& $4.4\times
10^{-6}$\hskip.5cm [b]&\\

& \vspace{-0.18truecm}& \\
\hline
& \vspace{-0.18truecm}& \\

$  B^+\, \to K^{*\,+}\,J/\psi$ & {\rm input}& $(1.35\pm
0.10)\times 10^{-3}$
\cite{PDBook}\\

& \vspace{-0.18truecm}& \\
\hline
& \vspace{-0.18truecm}& \\

$  B^+\, \to K_1^+(1270)\,J/\psi$ & $0.89\times\,10^{-3}$&
$(1.8\pm 0.5)\times 10^{-3}$\, \cite{Abe:2001}\\

& \vspace{-0.18truecm}& \\
\hline
& \vspace{-0.18truecm}& \\

$  B^0\, \to K_1^0(1270)\,J/\psi$ & $0.89\times\,10^{-3}$& $(1.3\pm 0.5)\times 10^{-3}$\, \cite{Abe:2001}\\

& \vspace{-0.18truecm}& \\
\hline

& \vspace{-0.18truecm}& \\

& $1.4\times 10^{-5}$\hskip.5cm [a]&\\

$  B^{+}\, \to K^{+}_1(1400) J/\psi$&
& $< 5\times 10^{-4}$\,\,\cite{PDBook}\\

& $8.1\times
10^{-5}$\hskip.5cm [b]&\\

& \vspace{-0.18truecm}& \\

\hline
& \vspace{-0.18truecm}& \\

$  B^+\, \to  K^{*\,0} \pi^+$  & input &$\left(1.9^{+0.6}_{-0.8} \right)\times 10^{-5}$\cite{PDBook}\\

& \vspace{-0.18truecm}&\\
\hline
& \vspace{-0.18truecm}& \\

$  B^0\, \to  K^{*\,+} \pi^-$  &input &$\left(1.6^{+0.6}_{-0.5} \right)\times 10^{-5}$\,\cite{PDBook}\\

& \vspace{-0.18truecm}&\\
\hline
& \vspace{-0.18truecm}& \\

$  B^+\, \to  K^{0}_1(1270) \pi^+$  & $1.0\,\times 10^{-5}$&==\\

& \vspace{-0.18truecm}&\\
\hline

& \vspace{-0.18truecm}& \\
$  B^0\, \to  K^{+}_1(1270) \pi^-$& $0.85\,\times 10^{-5}$ &==\\

& \vspace{-0.18truecm}& \\
\hline

& \vspace{-0.18truecm}& \\

$  B^+\, \to  K^{0}_1(1400) \pi^+$  & $0.54\,\times 10^{-5}$&$<\,2.6\,\times 10^{-4}$\, \cite{PDBook}\\

& \vspace{-0.18truecm}&\\
\hline

& \vspace{-0.18truecm}& \\
$  B^0\, \to  K^{+}_1(1400) \pi^-$& $0.46\,\times 10^{-5}$&$<\,1.1\,\times 10^{-3}$\, \cite{PDBook}\\

%%%%%%%%%%%%
  \hline
%%%%%%%%%%%%
\end{tabular}
\end{center}
\end{footnotesize}
\end{table}

%\bibliographystyle{apsrev}
%\bibliography{kappa1}

\end{document}